\documentclass[prl,twocolumn,showpacs,preprintnumbers,amsmath,amssymb]{revtex4}

\usepackage[dvips]{graphics}
\usepackage{tikz}
\begin{document}

\title{Unpredictability is perfectly possible in a deterministic universe}

\author{Chiara Marletto and Vlatko Vedral}
\affiliation{Clarendon Laboratory, University of Oxford, Parks Road, Oxford OX1 3PU, United Kingdom}

\date{\today}

\begin{abstract}
We revisit the vexed question of how unpredictability can arise in a deterministic universe, focusing on unitary quantum theory. We discuss why quantum unpredictability is irrelevant for the possibility of what some people call `free-will', and why existing `free-will' arguments are themselves irrelevant to argue for or against a physical theory. 
\noindent \end{abstract}


\maketitle                           %display desi d

The dynamical laws of unitary quantum theory (i.e., the Schr\"odinger equation, or the Heisenberg equation, or their relativistic generalisations) are deterministic.
Such determinism is often used to argue against unitary quantum theory. One of the most popular lines of argument is that determinisim allegedly removes the possibility of `free will'; while (the argument goes) a stochastic universe is perfectly compatible with free will (see e.g. \cite{Gisin}). This is perhaps the most extravagant objection to unitary quantum theory given that we do not have a physical theory of `free will' itself. The same argument is also used to call for modifying quantum theory's unitary laws, to make them explicitly stochastic, like in collapse-like theories, \cite{Bassi}. 

The free will argument is flawed at many different levels. First, it is a misconception that in unitary quantum theory everything is predictable. Yes, unitary quantum theory has a deterministic law of motion describing how $q$-numbered valued quantities evolve (these can be the quantum state of the universe, or the collection of its quantum observables, depending on the picture one is using), \cite{DEU, Ved}. These quantities are not directly measurable in a `single shot' way. In fact, once a unitary model of measurement is set up, it turns out that observed outcomes of given measurements are impossible to predict. This property, which we prefer to call `unpredictability' to distinguish it from randomness, is in fact a necessary condition for the appearance of randomness in unitary quantum theory and even in more general deterministic theories with the same structure as quantum theory, \cite{Mar}. 

The unpredictability of measurement outcomes is completely consistent with what is observed in routine experiments in quantum laboratories: when measuring an observable of a qubit prepared in a superposition of its eigenstates, it is impossible to predict which outcome will be observed. This is true even if one maintains that the qubit and the measuring apparatus evolve unitarily in a situation where the qubit and the measuring apparatus are entangled. That both outcomes occur (a factual property when one sticks to the fully unitary picture) is compatible with the impossibility of predicting the outcome (a counterfactual property, defined relative to the state of the measuring apparatus). 

So, even if one thinks that free will requires unpredictability as a necessary feature of physical laws (which incidentally we believe is untrue), unpredictability is perfectly possible in unitary quantum theory, in the same way as it is possible in a stochastic modification of quantum theory itself. 

Second, the concept of free will is vague and ill-defined - so it is a shaky basis to build a general argument against a given physical theory. Of course we all experience (and enjoy) the feeling of making our decisions spontaneously and autonomously, and it is comforting to know that this feeling is not in contradiction with our most fundamental understanding of the universe; but in order to understand exactly how physical laws allow for free will (or consciousness, or creativity -- call it whatever you like), one needs a physical theory of it, which we currently do not have. 

Another argument against unitary quantum theory is that its only available interpretation is the so-called ``Many-world" interpretation (see e.g. \cite{Wallace} for a comprehensive review). Typically one takes a specific version of the Many-World interpretation and argues against it. This often degenerates in a self-referential discussion about interpretations of interpretations, which moves away from the physical content of the theory. For instance, a popular argument is to use Occam's Razor to either say that Many Worlds is ruled out as it has too many worlds, or that it is the best interpretation because it has the fewest numbers of axioms. We believe that both arguments are bad, because the `simplicity' in Occam's Razor cannot be objectively quantified in the first place, and hence it is a misleading criterion to evaluate if a theory is a good explanation. 

Unitary quantum theory is consistent; it provides a good explanation of all the experimental observations so far, and (unlike some of its stochastic variants) it is also compatible with properties of general relativity, such as locality and the equivalence principle. It will no doubt have to undergo changes to accommodate gravity exactly, but we bet that unpredictability and determinism are here to stay, together with other phenomena like superpositions and entanglement. It is therefore high time we took all these aspects of unitary quantum physics seriously \cite{Deutsch2}.

\textit{Acknowledgments}: The Authors thank the Moore Foundation, the John Templeton Foundation, and the Eutopia Foundation.

\end{document}